\documentclass[runningheads]{llncs}
\usepackage[T1]{fontenc}
\usepackage{graphicx}

\usepackage{todonotes}
\usepackage{hyperref}
\usepackage{color}

\begin{document}
\title{KamerRaad: Enhancing Information Retrieval in Belgian National Politics through Hierarchical Summarization and Conversational Interfaces}
%
%
\author{Alexander Rogiers\orcidID{0000-0003-4241-0596} \and
Maarten Buyl\orcidID{0000-0002-5434-2386} \and\\
Bo Kang \orcidID{ 0000-0002-9895-9927} \and
Tijl De Bie\orcidID{0000-0002-2692-7504}}

\authorrunning{A. Rogiers et al.}
\titlerunning{KamerRaad} 
%
\institute{Ghent University, Ghent, Belgium\\
\texttt{\{alexander.rogiers, maarten.buyl, bo.kang, tijl.debie\}@ugent.be}}

\maketitle 
\begin{abstract}
KamerRaad is an AI tool that leverages large language models to help citizens interactively engage with Belgian political information. 
The tool extracts and concisely summarizes key excerpts from parliamentary proceedings, followed by the potential for interaction based on generative AI that allows users to steadily build up their understanding.
KamerRaad's front-end, built with Streamlit, facilitates easy interaction, while the back-end employs open-source models for text embedding and generation to ensure accurate and relevant responses. 
By collecting feedback, we intend to enhance the relevancy of our source retrieval and the quality of our summarization, thereby enriching the user experience with a focus on source-driven dialogue.

\end{abstract}

\section{Introduction}
Fundamental for democratic participation is that politicians' views, policies, and actions are accessible to citizens. 
Such critical insights are contained in publicly available documents, yet they wildly vary in format and are dense with specialized terminology. 
The dispersion of this information across documents from plenary and committee meetings on different days and various types (discussions between politicians vs. proposals of resolutions and laws) further complicates their retrieval and contextualization.
Passing the sources to Large Language Models (LLMs) is a potential solution, but the source documents are too long for even one to fit into the context window of these models.

\begin{figure}[t]
    \centering
    \includegraphics[width=0.75\textwidth]{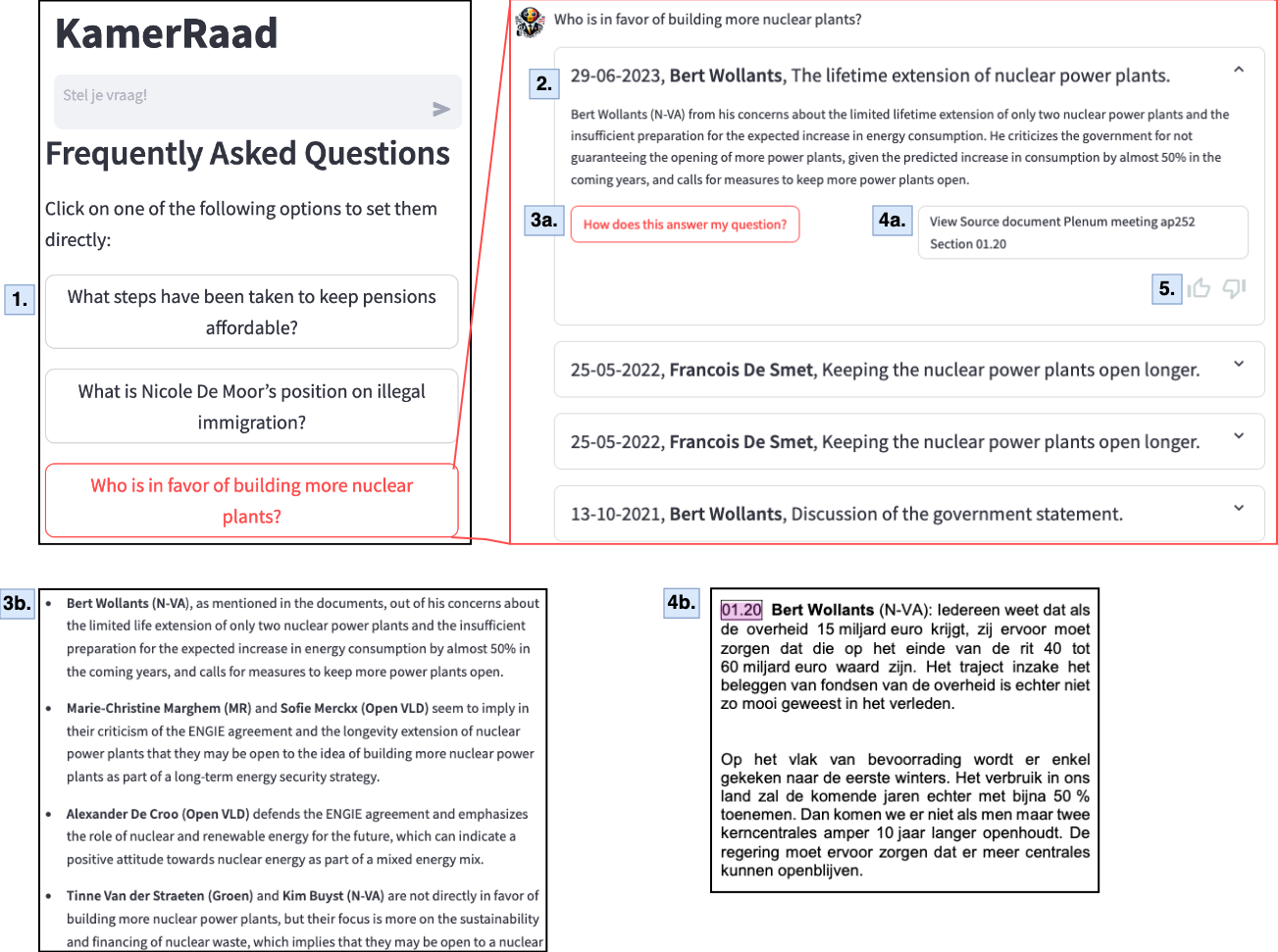}
    \caption{KamerRaad UI displaying the entire user flow for the example question "Who's in favor of building more nuclear power plants?". Starting from \textbf{1.} a query input field with suggested questions to \textbf{2.} relevant summaries with the options to \textbf{3.} generate a response that clarifies how the source answers the question, \textbf{4.} view the complete source document and \textbf{5.} give explicit about the retrieval and generation. The text in this figure has been translated to English to be accessible to the reader.}
    \label{fig:KamerRaad_user_flow}
\end{figure}

We propose KamerRaad, an AI tool that uses \emph{hierarchical summarization} to manage the context length while maintaining the source's integrity.
This enhances the accessibility of the sources by enabling interaction in a conversational tone where users start with a question and are guided through summaries to detailed responses, exemplified in Figure \ref{fig:KamerRaad_user_flow}.
KamerRaad, developed with the Belgian elections in mind, is freely available at \href{https://kamerraad.org}{kamerraad.org}. 
A video demo is available on \href{https://youtu.be/Tl7aicZsk_Y}{youtube}.

Our methodology is grounded in the principles of \emph{Retrieval-Augmented Generation} (RAG), a technique that supplements language model inputs with information from external databases\cite{lewisRetrievalAugmentedGenerationKnowledgeIntensive2021}. However, to apply RAG, substantial challenges had to be overcome, to meet context length limitations, and to ensure responsiveness, transparency, factuality, and awareness of contextual information such as temporality and entities relevant to the retrieved documents.

\subsubsection{Contributions} By addressing the limitations of LLMs in processing long documents, KamerRaad manages the challenge posed by extensive parliamentary records, ensuring that users receive precise and contextually relevant information without overwhelming the model's processing capabilities. Beyond the standard RAG approach, KamerRaad introduces the pre-processing step of hierarchical summarization to manage the context length of the source documents for generation while also  extracting metadata tags to improve retrieval.

\section{Use Cases}
KamerRaad's structured conversational interface maintains a direct link to source documents, aiding both expert policymakers and the general public in navigating this complex corpus of political information. 
For expert users, such as parliamentary aides and policy analysts, KamerRaad offers a mechanism to trace summaries and generated responses back to its source documents, enhancing the reliability of their analyses. 
For non-expert users, it simplifies the complex landscape of national policy by enabling a contextual understanding through its generative and interactive features. 



\begin{figure}
    \centering
    \includegraphics[width=0.9\textwidth]{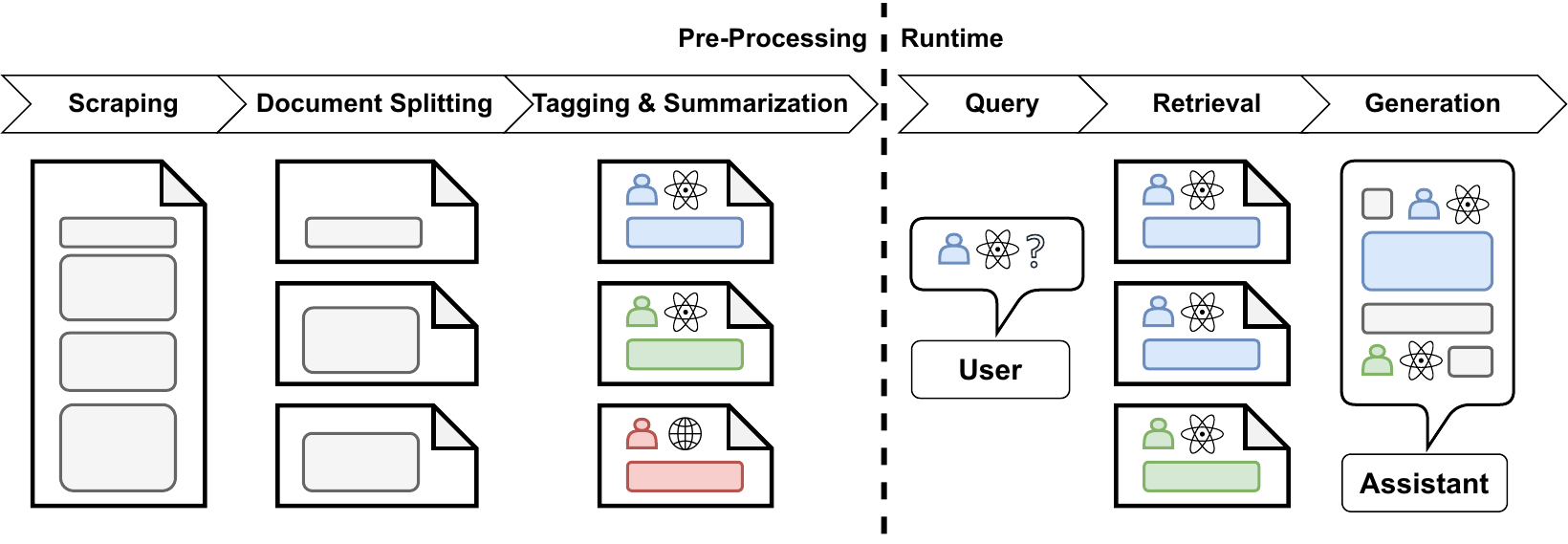}
    \caption{KamerRaad processing pipeline. During pre-processing we scrape and chunk raw documents. During tagging and summarization we enhance each chunk with a full summary, a short summary, politician and topic. This metadata is represented as colors for the politicians and symbols for the topic. At runtime the user is presented first with document summaries relevant to their query. Relevance is calculated as cosine similarity between the prompt embedding and the summary embedding. The generative model provides a response when the user shows interest in the sources by interacting with the UI. A direct link to the source chunks is always maintained in the response as visualized by the colors and topic reoccurring in the speech bubble.}
    \label{fig:KamerRaad_process_flow}
\end{figure}

\section{Design Choices and Implementation}
KamerRaad transforms documents from the Belgian national parliament into a searchable database by tagging and categorizing parliamentary discussions available at \href{https://www.dekamer.be}{www.dekamer.be}. 
Each politician's statement is segmented into manageable chunks, which are then summarized to address a key challenge in RAG systems: managing the limited context window of language models.

From these chunks we generate both comprehensive and concise, one-line summaries forming our hierarchical summarization approach.
Comprehensive summaries ensure that detailed, relevant information is included in the response generation context, while one-line summaries provide a high-level overview, allowing multiple summaries to fit within a single LLM prompt's context window. 
This approach optimizes the content for LLMs, ensuring that the retrieval system fetches information that is both pertinent and concise enough to be processed with off-the-shelf LLMs.

Metadata tags for politicians, political parties, and topics further enhance retrieval efficiency. 
These tags allow users to filter information not only based on document relevance—which is defined by the cosine similarity between the user prompt and chunk embeddings—but also on specific political figures and themes. 
Users can then interact with these sources, prompting KamerRaad to generate more comprehensive responses based on the refined input.

\subsubsection{Implementation Details}
Pre-processing involved the development of custom scraping and text splitting tools. 
The summarization and tagging process leveraged GPT-4 \cite{openaiGPT4TechnicalReport2024}, chosen because of its popularity at the time of writing.

During the inference phase, KamerRaad utilizes open-source models for both document section retrieval and user response generation. Specifically, the retrieval process is handled by \texttt{BAAI/bge-m3} \cite{bge-m3}, while response generation utilizes \texttt{GEITje-7B-ultra}\footnote{\href{https://huggingface.co/BramVanroy/GEITje-7B-ultra}{huggingface.co/BramVanroy/GEITje-7B-ultra}}. 
Figure \ref{fig:KamerRaad_process_flow} shows an overview of the pre-processing and runtime data flow. 

KamerRaad's user interface, built with Streamlit\cite{streamlit2024} and hosted on Google Cloud, offers scalable, real-time interactions. 
Users can provide explicit, binary feedback within the interface on the responses, which will be used to improve the system's retrieval model and summarization tactics.

\section{Conclusion}
This paper highlights KamerRaad's innovations in political information retrieval, including hierarchical summarization for extensive documents and the use of large language models to clarify complex topics. 
By staging user interactions from questions to sourced answers, KamerRaad promotes a deeper understanding of political discourse and actively supports democratic engagement.
\begin{credits}
\subsubsection{\ackname} The research leading to these results has received funding from the Special Research Fund (BOF) of Ghent University (BOF20/IBF/117), from the Flemish Government under the ``Onderzoeksprogramma Artificiële Intelligentie Vlaanderen'' programme, and from the FWO (project no. G0F9816N, 3G042220, G073924N).
\subsubsection{\discintname}
The authors have no competing interests to declare that are relevant to the content of this article.
\end{credits}

%
%
%
%

\bibliographystyle{splncs04}
\bibliography{references}
\end{document}